\documentstyle[twocolumn,psfig]{mnv2}

\newif\ifAMStwofonts
\AMStwofontstrue

\def\kms{\rm{km s^{-1}}}
\def\mpc{\rm{h^{-1}Mpc}}
\def\cA{${\cal A}$}
\def\calH{{$\cal H$}\,}
\def\gsim{~\rlap{$>$}{\lower 1.0ex\hbox{$\sim$}}}
\def\dhi{n_{_{\rm HI}}}
\def\dhhi{\hat n_{_{\rm HI}}}
\def\nhis{n^s_{_{\rm HI}}}
\def\nhi{\dhi}
\def\dhij{(n_{{\rm HI}})_{j}}
\def\dhii{(n_{{\rm HI}})_{i}}

\def\vp{v_{\rm p}}

\newcommand{\norm}{($\Omega_{\rm bar}h^2)^2 \,J^{-1}\,H(z)^{-1}$}

\def\ltsim{\lower.5ex\hbox{$\; \buildrel < \over \sim \;$}}
\def\gtsim{\lower.5ex\hbox{$\; \buildrel > \over \sim \;$}}
\def\ltsim{\lower.5ex\hbox{$\; \buildrel < \over \sim \;$}}
\def\gtsim{\lower.5ex\hbox{$\; \buildrel > \over \sim \;$}}
\def\th{\theta}
\def\tht{\tilde \theta}
\def\cv{ v}
\def\cvt{ \tilde v}
\def\wt{ \tilde w}
\def\ut{ \tilde u}

\def\vx{{\bf x}}
\def\vk{{\bf k}}
\def\vr{{\bf r}}
\def\kpar{k_\parallel}
\def\kper{\vk_\perp}
\def\dtl{{\tilde {\delta}}^{\rm los}}
\def\dt{{ \tilde {\delta}}}
\def\dl{{ {\delta}}^{\rm los}}
\def\d{{ {\delta}}}

\def\vv{{\bf v}}

\def\cm{\,{\rm cm}}

\def\s{\,{\rm s}}
\def\K{\,{\rm K}}
\def\kms{\mbox{km\,s$^{-1}$}}

\def\dd{\,{\rm d}}
\def\dhik{(n_{{_{\rm HI}}})_{k}}

\newcommand{\op}{Ly$\alpha$\ }
\newcommand{\hi}{\mbox{H{\scriptsize I}}}

\begin{document}
\title[]{A first step towards a direct inversion of  the Lyman forest
in QSO spectra}

\author[Nusser \& Haehnelt] { Adi Nusser$^1$ and Martin Haehnelt$^2$ \\
$^1$Max-Planck-Institut f\"ur Astrophysik, Karl-Schwarzschild-Str. 1,
D--85740 Garching b. M\"unchen, Germany\\ 
$^2$ Institute of Astronomy, Madingley Rd, CB3-0HA, Cambridge, UK}

\maketitle

\begin{abstract}

A method for the recovery of the real space line-of-sight mass density 
field from Lyman absorption in QSO spectra is presented. 
The method makes use of a Lucy-type algorithm for the recovery 
of the \hi\ density. The matter density is inferred 
from the \hi\ density assuming that the absorption is due to a
photoionized  intergalactic  medium which traces the mass distribution 
as suggested by recent numerical simulations. Redshift distortions are 
corrected iteratively from a  simultaneous estimate of the peculiar
velocity. The method is tested with mock spectra obtained from N-body 
simulations. The density field is recovered reasonably well up to
densities where the absorption features become strongly saturated. 
The method is an excellent tool to study the density probability
distribution and clustering properties of the mass density in the
(mildly) non-linear regime.  Combined with redshift surveys along 
QSO sightlines the method will make it possible to relate the 
clustering of high-redshift galaxies  to the clustering of 
the underlying mass density. We further show that accurate estimates 
for \norm \ and higher order  moments of the density probability function 
can be obtained despite the missing high density tail of the density
distribution if a parametric form for the probability distribution 
of the mass density is assumed. 

\end{abstract}


\section{Introduction}

The Lyman forest in QSO absorption spectra is now generally believed
to be due to absorption by large-scale neutral hydrogen (\hi) density 
fluctuations of moderate amplitude in a warm photoionized
intergalactic medium (IGM). 
This relatively new paradigm for the \op forest differs considerably 
from the conventional ``cloud'' picture which had been advocated for
two decades and in which the Lyman forest is due to a superposition
from discrete absorbers with small cross sections (see Rauch 1998 for 
a review). The new picture picture had been investigated
using analytical calculations (e.g. Bond, Szalay \& Silk;
Mc Gill 1990; Bi, B\"orner \& Chu 1992) but was 
only generally accepted after the coherence length of the 
absorbing structure could be measured accurately and turned out to be 
of order several hundred kpc (Dinshaw et al. 1994, e.g. Bechthold et al. 
1994). The picture was then further sustained by 
(hydrodynamical) cosmological simulations of gas in dark-matter 
dominated universes (Cen et
al.~1994; Petitjean, M\"ucket \& Kates 1995; Zhang, Anninos \& Norman
1995; Hernquist et al.~1996; Miralda-Escud\'e et al. 1996). 
Important results of these simulations are a tight correlation
between the \hi\ and the dark matter  distribution (on scales larger 
then the Jeans length of the IGM) and a simple temperature-density
relation for the IGM which depends only on the reionization history of
the universe (e.g. Hui \& Gnedin 1997, Haehnelt \& Steinmetz 1998).

As a consequence of this new picture the Lyman forest can be used to
probe the distribution and clustering of dark matter at high
redshift. For example, Bi \& Davidsen (1997) have developed a simple
analytic model for the IGM to generate artificial QSO absorption
spectra for a variant of the cold dark matter (CDM) cosmogony. They
were able to reproduce the characteristic properties of observed
absorption spectra. Croft et al. (1998) showed that absorption spectra
provide important information on the shape and the amplitude of
the power spectrum of mass fluctuations. Croft et al. used the
following procedure. They first obtained a linearized flux
distribution by applying the Gaussianization scheme proposed by Weinberg
(1992). They then inferred the shape of the linear power spectrum of
dark matter density fluctuation from the power spectrum of the
linearized flux and used mock spectra from numerical simulations to
determine the amplitude of the power spectrum. Gnedin \& Hui (1998) used 
mock spectra generated from simulations of collisionless particles 
run with a particle-Mesh code modified to mimick pressure effects 
of the gas to investigate the effect of amplitude and the power spectrum of
dark-matter  fluctuations on the  column density distribution of \op 
absorption systems.  

In this paper a complementary approach is taken.  We propose to use an
analytical model of the IGM for a direct inversion of the absorption
features in QSO spectra. 
This approach has the advantage that no particular cosmological model 
has to be assumed. Furthermore,  the actual real space density along 
the LOS and its probability distribution can be studied and 
a direct link to other observations is posssible.

The paper is organized as follows. 
In Section 2  we describe the assumed model for the Lyman 
absorption. Section 3  presents the inversion algorithm for recovering the 
{\it real} space dark matter (DM) density along the l.o.s.,
concentrating on the \op forest. In section 4 we test the inversion 
procedure with mock spectra generated from numerical simulations of 
collisional dark matter and show how to estimate  higher order moments 
of the density probability distribution  and \norm\  (the 
parameter combination of baryonic density $\Omega_{\rm bar}h^{2}$, 
ionizing flux $J$ and Hubble constant $H(z)$  which determines the
mean flux level in QSO absorption spectra). In section 5 we discuss 
possible applications and give our conclusions.

\section{Lyman absorption by a photoionized intergalactic medium}

Absorption spectra are normally presented in the form 
of a normalized flux which (neglecting noise and instrumental 
broadening) can be related to the optical 
depth as 
\begin{equation}
F(w)= I_{\rm obs}(w)/I_{\rm cont}(w) =  {\rm e}^{-\tau(w)} , 
\end{equation}
where $\tau$ is the optical depth, $w$ is the  redshift space
coordinate, $I_{\rm obs}(w)$ is the observed flux and $I_{\rm
cont}(w)$ is the flux emitted from the quasar which would be 
observed without intervening absorption and has to be estimated 
from the data as well.

The optical depth in redshift space due to  resonant  \op scattering 
is related to the neutral hydrogen density, $\dhi$, along the
line-of-sight (l.o.s) in real space as (Gunn \& Peterson 1965, Bahcall 
\& Salpeter 1965)  
\begin{equation}
\tau(w)=  \sigma_{0} \;\frac{c}{H(z)}\; \int_{-\infty}^{\infty} 
\dhi (x)\;  {\cal H} [w-x-\vp (x), b(x)]  \dd x,
\label{tau}
\end{equation}
where $\sigma_{0}$ is the effective cross
section for resonant line scattering,  $H(z)$ is the Hubble constant at
redshift $z$,  $x$ is   the real space coordinate (in $\kms$), 
\calH is the  Voigt profile normalized such that $\int{\cal H }\, \dd x =1 $, 
$v_{\rm p}(x)$ is the l.o.s.  peculiar velocity, and $b(x)$ is the Doppler 
parameter due to   thermal/turbulent broadening. 
For moderate optical depths the Voigt profile  is well approximated 
by a Gaussian,
${\cal H }  =\frac{1}{\sqrt{\pi}\, b}\,\exp{[-(w-x-\vp (x))^2/b^2)}$. 
Assuming that hydrogen is  highly ionized and in 
photoionization equilibrium ($\nhi  \propto    n_{H}^2   J^{-1} $) 
we  have 
\begin{equation}
\dhi  = \dhhi \left(\frac{n_{\rm H} (\vx)}{{\bar n_{\rm H}}}\right)^\alpha 
      =  \dhhi \left(\frac{\rho \vx)}{{\bar \rho}}\right)^\alpha, 
\label{nhi}
\end{equation}
where $\dhhi$  is the neutral hydrogen density at mean total gas 
density and  $1.56<\alpha <2$ (Hui\& Gnedin 1997).    
The second relation assumes  that the
gas density traces the dark matter density.  
At the low densities considered here shock heating is not important 
ant the gas is at the photoionization equilibrium temperature.
The Doppler parameter can then be related to  
the total gas  density and temperature as 
\begin{equation}
b(x)= 13 \left( \frac{\hat T}{10^{4}\K} \right )^{0.5}  
\left(\frac{\rho_{\rm DM} (\vx)}{{\bar \rho_{\rm DM}}}\right)^\beta 
\kms,
\label{doppler}
\end{equation}
where  $\hat T$ is the temperature at the mean density
and $0 <\beta < 0.31$ (Hui \& Gnedin 1997). 

Combining equation (2), (3) and (4) we get 
\begin{equation}
\tau(w)=  {\cal A} (z)  \int_{-\infty}^{\infty}  
\left(\frac{\rho_{\rm DM}(\vx)}{{\bar \rho_{\rm DM}}}\right)^\alpha
{\cal H} [w-x-\vp (x), b(x)] \dd x
\end{equation}
\begin{eqnarray}
{\cal A} (z)&=& \sigma_{0}  \frac{c}{H(z)} \; \dhhi \nonumber \\
&\sim &  0.12 h^{-1} \; \Omega_{\rm mat}^{-0.5}\;
\left (\frac{\Omega_{\rm bar}h^2}{0.0125}\right )^{2} \;\times \nonumber \\
&&\left ( \frac{\Gamma}{10^{-12}\s^{-1}} \right)^{-1}\; 
\left ( \frac{\hat T}{10^{4}\K} \right )^{-0.7}  \,
\left ( \frac{1+z}{4} \right )^{4.5} .\nonumber \\ 
\label{tauf}
\end{eqnarray}
where $\Omega_{\rm bar}$ and $\Omega_{\rm mat}$ are  
the baryonic and total matter density in terms of the critical 
density and $\Gamma$ is the photoionization rate per hydrogen 
atom ($\Gamma$ is  related to the flux of ionizing radiation $J_{\nu}$
as $\Gamma = 4\pi\, \int{ \dd \nu \, \sigma_{\nu}\, J_{\nu}/h\nu}$, 
where $\sigma_{\nu}$ is the hydrogen  absorption cross section).    
To obtain the second relation in equation (6) we have 
 have taken $\alpha_{\rm rec} = 4.7 \times  10^{-13} (T/10^{4}
\K)^{-0.7} \cm^{-3} \s^{-1}$ for the hydrogen recombination
coefficient,  $\sigma_{0} =4.5 \times 10^{-18}\cm^{2}$  as the
effective cross section for resonant \op scattering and used 
the  high redshift approximation for the Hubble constant, 
$H(z) = H_{0}\, \Omega_{\rm mat}^{1/2}\, (1+z)^{3/2}$.   

\section{The inversion algorithm}

\subsection{The basic iterative scheme} 

The proposed scheme for recovering the l.o.s. density from the flux 
is motivated by Lucy's method (Lucy 1974). In order to demonstrate the
method let us,
for the time being, consider the case where $\tau \ll 1$
and $\vp =0$.
In this case, we have
\begin{equation}
1-F(w)= \int\dhi(x)G(w,x) dx .
\end{equation}
This is a linear integral equation with a positive Kernel
$G(w,x) = {\cal H} [w-x, b(x))] \sigma_{0}\, c/H(z)$
and can be solved for $\dhi$ using Lucy's
iterative method. This is done in the following way. Divide $w$ into
equal bins of size $\Delta w$ (the data are actually given in bins of
$w$ anyways) and set $f_i = 1-F_i$. Provide an initial guess for $\dhij$,
say $\dhij^{0}= 1-F_j \,H(z)/(\sigma_0\,c)$. 
Denote the values of $\dhij$ at the $r$th iteration by 
$\dhij^r$ and evaluate the sum 
\begin{equation}
f^r_i =\sum_j \dhij^r \;G_{ij}^{r}\; \Delta w \label{iterf},
\end{equation}
where
\begin{equation}
G_{ij}^{r}= {\cal H}[w_{i} - x_{j},b_{j}^{r}] \;\frac{\sigma_{0} \; c}{H(z)}.
\end{equation}
The  $(r+1)-$th  estimate of $\dhij$ is then
\begin{equation}
\dhii^{r+1}=\left[\frac{1}{2m+1}\sum_{k=i-m}^{i+m}\!\dhik^r\right]\; \frac{
\sum_j (f_j/f^r_j)\tilde G_{ij}}
{\sum_j \tilde G_{ij}} \label{iter} ,
\end{equation}
where $\tilde G$ is some Kernel which in principle may be chosen to 
be different from $G$. Lucy's method, however,  uses $\tilde G=G$.
The averaging over $2m+1$ adjacent bins mitigates the effect 
of noise in the data. We work with $m=3$. 

Now we generalise the iteration scheme to the case where $\tau$ is
not necessarily small. In this case, 
the relation between $F$ and $\tau$ is nonlinear. However, since the relation
is monotonic, we can still use a modified version of equation 
(\ref{iter}), 
\begin{equation}
f^{r}_i=1-\exp[-\sum_j \dhij^r\; G_{ij}^r \;\Delta w] \label{iterfnl} .
\end{equation}

How do we decide when to stop the iterations?
Suppose that in the $j-$th bin, the 1-$\sigma$ error in the measurement
of $F_j$ is $\sigma_j$. 
Let us define the quantity
\begin{equation}
\chi^2=\sum_j\frac{1}{\sigma_j^2}\left(f_j - f^r_j \right)^2 \label{stop} .
\end{equation}
We choose to stop  at the $r-$th iteration
when the value of $\chi^2$  drops below the number of data bins. 
It implies that the observed flux is consistent with being a 
noisy realisation of the reconstructed flux.

\subsection{Recovery of the matter density in redshift space}

We write (\ref{tau}) in terms of 
the redshift coordinate $s\equiv x+\vp (x)$, instead of $x$. If
the coordinates $x$ and $s$ are related by a
one-to-one mapping, then  the velocity 
and the real space coordinate,
can be expressed uniquely as functions  of the redshift coordinate $s$.
Working with $s$, equation (\ref{tau}) takes the following form
\begin{eqnarray}
\tau(w)&=& \sigma_{0}\; \frac{c}{H(z)}\; \int_{-\infty}^{\infty} 
\nhi\left[s-\vp(s)\right]\left[1-\frac{\dd
\vp(s)}{\dd s}\right]  \; \nonumber \\
&&\nonumber \\
&&  {\cal H} [w-s, b(s-\vp(s))] \dd s \nonumber \\
\label{taus}
\end{eqnarray}
For brevity, we have maintained the same notation
for the peculiar velocity as a function of $s$.
According to the continuity equation, the density in redshift
space, $\nhis$, is  related to the density in real space by
\begin{equation}
\nhis(s)=\nhi[s-\vp(s)] \left[1-\frac{\dd
\vp(s)}{\dd s}\right] . \label{cont}
\end{equation}

Therefore (\ref{taus}) becomes
\begin{equation}
\tau(w)=\int_{-\infty}^{\infty} K(w,s) \; \nhis(s) \; \dd s 
\end{equation}
where $K(w,s)={\cal H} [w-s, b(s-\vp(s))]
\sigma_{0}\, c/H(z)$ and the iterative scheme described in the last 
section can be applied to obtain the  redshift space density 
along the l.o.s.  The parameter $b$ depends on $(s-\vp)$ implicitly 
via $\nhi$. Because $b$ is a weak function of $\nhi$, we assume that 
$b$ is constant in the reconstruction algorithm. This will greatly 
simplify the application of the algorithm. Tests of the algorithm, 
however, are done using mock spectra generated 
with varying $b$ according to equation (\ref{doppler}).

To relate the \hi \ density to the matter density we use 
equation (\ref{nhi} and (\ref{tauf}). We still need to know the value
of \cA\ to obtain $\rho_{DM}/\bar  \rho_{\rm DM}$.  \cA\ has 
been estimated by adjusting the mean flux of  mock spectra generated 
from hydrodynamical simulations to match the mean
flux of  observed  QSO absorption spectra (Rauch et al.~1997).  
The  accuracy and possible biases of this determination 
have  not yet been investigated in detail (see also Weinberg et
al. 1997), but we consider the value of ${\cal A}$ ($\propto$ \norm ) 
to be presently known with about 30 percent accuracy. For most of 
the paper we therefore assume that we know the value of \cA\ .
In section 4.3 we show how the  
value of $\cal A$ ($\propto$ \norm )  can be estimated directly from 
the distribution of the recovered  quantity ${\cal A}^{1/\alpha}  
(\rho_{DM}/\bar  \rho_{\rm DM})$  by assuming a parametric form for 
the probability distribution of the matter density.

\subsection{From redshift space to real space}

The correction for redshift distortions requires knowledge of the 
velocity field.  The 3-dimensional velocity and mass density 
fluctuations are tightly related. However, non-linear velocity-density 
relations which are easy to implement generally relate real space
quantities, as does the relation (\ref{nhi}) between the \hi\ and
matter density.  We have therefore  to resort to an iterative 
procedure to derive the real space matter and \hi\ densities from the 
estimated redshift space \hi\ density. 
Two issues need to be stressed here: (i) the velocity field is not
uniquely specified by the l.o.s. density field but influenced by the
unknown 3D matter distribution, and (ii) redshift distortions 
can and do result in  multi-valued zones where regions which do not 
overlap in real space are mapped onto the
same redshift coordinate. In appendix A  we describe a method 
which resembles Wiener filtering and addresses the first point. It allows us 
to obtain the most probable velocity field which is consistent 
with the estimated l.o.s. DM density field  and  has the statistical 
properties of a gravitationally clustering Gaussian random field 
of a given power spectrum. In principle  the power spectrum can also 
be determined from the  recovered density field but for simplicity 
we have chosen  to adopt a power spectrum a priori. The results do not change
much for reasonable choices of  the assumed power spectrum. 
Multi-valued zones mainly occur in  regions of high density 
where the inversion is difficult due to saturation effects 
in the spectrum. We did not try to correct for  this effect, 
but discuss some of the biases introduced by peculiar velocities 
in section 4.    

The iterative  scheme to correct  for redshift distortions
which we have adopted can be summarized as follows:

\begin{enumerate} 
\item 
Assume that the redshift-space and real-space 
\hi\  density are equal and use 
(\ref{nhi}) to compute  the  matter density field along the l.o.s. 
\item
Estimate the peculiar velocity along the l.o.s. 
using the method described in the appendix. 
\item 
Use the estimated  peculiar velocity field to correct for redshift  
distortions in the  density field. 
\item 
Use the corrected density field to obtain 
a better estimate for the peculiar velocity field.
\item 
Repeat steps (ii)-(iv) until convergence is achieved.  
\end{enumerate} 

The scheme proved to be efficient; it typically converges after a few 
iterations.  Iterative schemes of this kind have been used for
correcting redshift distortions in galaxy redshift survey 
(e.g. Yahil et. al. 1991).

\section{Tests with numerical simulations}

Ideally one would like to test the method with a large 
high-resolution hydrodynamical simulation of gas and 
dark matter.  However, simulations of collisionless particles 
are still vastly superior in dynamical range and  
especially speed. We have therefore chosen
to test our inversion algorithm  with a high resolution
N-body simulation of pure collisionless dark matter particles
where we used the relations in section 2.1 to obtain the  
\hi\ distribution  from the smoothed dark matter density 
field. As demonstrated e.g. by Gnedin (1998) such a procedure 
results in a realistic \hi\  distribution (apart from the high-density 
regions which are not of interest here).  

The simulation used was run on the Cray T3E parallel
supercomputer in Garching with a modified version (MacFarland et. al.
1997) of Couchman's ${\rm P}^3{\rm M}$ code (Couchman et. al. 1995). The
initial conditions were generated from the power spectrum for a
standard cold dark matter (CDM) universe with  $\Omega =1$ and 
$H_0=50{\rm km/s}$
cubic box of comoving length of $60 {\rm h^{-1} Mpc}$.  The simulation
was normalised such that the linear $rms$ density fluctuations in a 
sphere of $800{\rm km ~s^{-1}}$ was 0.5 at  redshift $z=0$. 
The softening parameter was $13.2\%$ of the mean 
particle separation and the mesh size was $N=512$ in one dimension.

\begin{figure}
\centering
\mbox{\psfig{figure=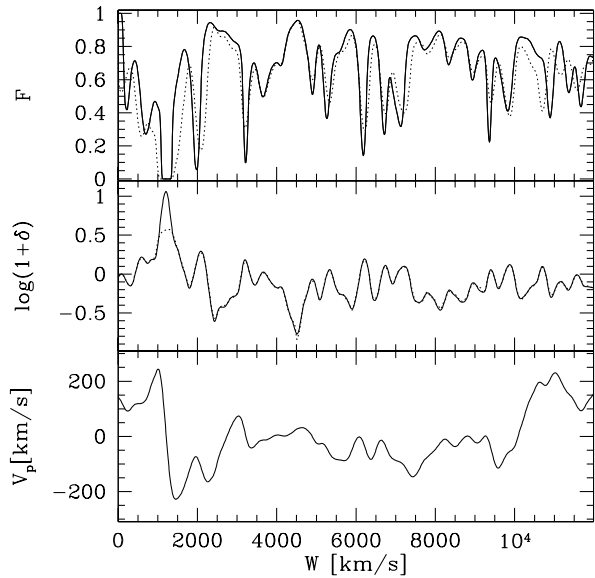,height=3.1in}}
\vspace{0.25cm}
\caption{ Normalized flux (top), density (middle) and velocity
(bottom) along one line-of-sight through the simulation box at $z=3$. 
The dotted curve in the top panel shows the flux with  peculiar 
velocities set to zero and the dotted curve in the middle panel 
shows the corresponding  recovered density field.}
\label{flux:sim}
\end{figure}

\subsection{Inverting mock spectra} 

We have generated mock spectra from velocity and density fields along
lines of sight randomly drawn from the simulation box. We used the
output of the simulation at two redshifts, $z=3$ and $z=0$, to
investigate the effect of varying the fluctuation amplitude of the
density field.  We emphasize here that  the mock spectra generated from the
simulation at $z=0$ are not meant to resemble  absorption spectra 
at $z=0$.  The density and velocity fields in the simulations 
show significant variations in structure and amplitude 
at the two redshifts and the $z=0$ spectra are investigated 
with the same mean flux as the  $z=3$ spectra in order 
to test the inversion  algorithm under  varying conditions.

\begin{figure}
\vspace{0.2cm}
\centering
\mbox{\psfig{figure=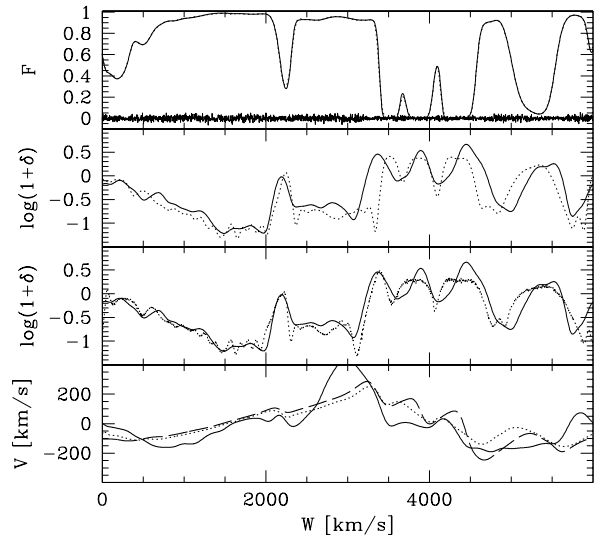,height=3.2in}}
\vspace{-0.7cm}
\caption{{\it Top panel:}  Normalized flux (solid curve) 
in a line-of-sight through the simulation  at $z=0$ and the flux 
corresponding to the density and peculiar
velocity  filed  recovered with a Lucy-type inversion algorithm
(dotted curve).  The noise  added to the input spectrum 
is shown separately at the bottom of the panel.  
{\it Second panel:} True real space density in the 
line-of-sight (solid curve) and recovered redshift-space matter
density (dotted curve).
{\it Third panel:} 
True real-space density  (solid curve, 
same as in second panel) and real-space matter density recovered 
with applying redshift corrections (dotted curve). 
{\it Bottom:panel } The true l.o.s. peculiar velocity field (solid
curve). The peculiar velocity estimated  from the recovered density 
 in redshift space (dotted curve). For comparison, also shown is the 
velocity recovered from the true real space density (dashed line). }
\label{los:sim}
\end{figure}

Reliable velocity and density fields cannot be derived from the simulation on
scales much smaller than the mean particle separation. We therefore
applied Gaussian smoothing of widths $0.3\mpc$ and $0.6\mpc$,
respectively, to the outputs of the simulation at $z=3$ and $z=0$.
Both smoothing scales correspond to the same physical scale
($60\kms$). The $rms$ values of the smoothed
density fluctuations were $ 0.89$ and $3.52$, respectively. 
The density of neutral hydrogen was assumed to follow the relation 
(\ref{nhi}) with $\alpha=1.7$. The absorption optical depth was
computed according to equation (\ref{tau}), (\ref{nhi}), 
and (\ref{doppler}) with $b_0=30\kms$.  For both redshifts the 
value of \cA\ was adjusted so that the mean normalized flux of a 
large ensemble of mock
spectra was 0.69 (about the  observed mean flux at $z=3$ as determined
by Rauch et al.~1997). The spectra
were convolved  with a Gaussian of 8km/s FWHM to mimick instrumental 
broadening and  photon/read-out noise was added  with a total S/N=50 
per 3km/s pixel.

The solid curve in the top panel of Fig.1 is a mock 
spectrum generated from the simulation output at $z=3$. The dotted
curve is the same spectrum with the  peculiar velocities set to zero. 
The peculiar motions do not only introduce a systematic  shift but 
also a narrowing of the absorption features (see also Weinberg et al.
1998  for a discussion of this point).  Furthermore, 
unsaturated lines get deeper due to the enhanced
clustering in redshift space. The mean flux is also affected.  For
constant \cA \ it increased from 0.66 to 0.69 including peculiar
motions while the rms flux fluctuations increased from 0.31 to 0.35.

\begin{figure}
\centering
\vspace{-3.5cm}
\mbox{\psfig{figure=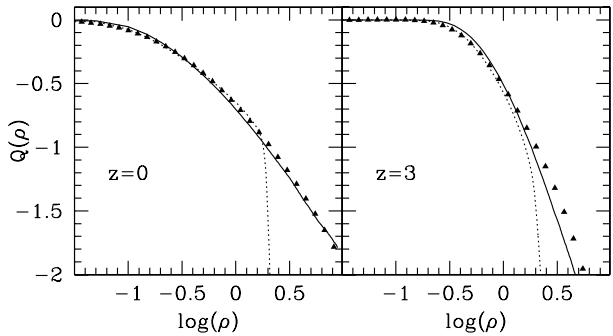,height=3.3in}}
\caption{The cumulative probability distribution of the true density
field (solid curve) and that of the density field recovered (dotted
curve) from mock spectra generated  from the simulation at $z=0$
(left) and $z=3$ (right). The triangles are the corresponding
parametric form derived from an Edgeworth expansion of the PDF
(\ref{edge}) as explained in the text.}
\label{logfit}
\end{figure}

Figure 2 shows some typical results of our inversion procedure. In the
top panel a mock spectrum is plotted (solid line) together 
with the spectrum corresponding to a recovered density and peculiar 
velocity field (dotted line). The smoothness of the dotted curve 
demonstrates that our inversion procedure  succeeds in mitigating 
the noise in the data (which for clarity is shown separately at the 
bottom of the panel).  

Let us now  have a look at the  DM density 
field  recovered from a mock spectrum generated with no peculiar
velocities as shown in the middle panel of Figure 1. In this case 
the correspondence is excellent up to densities 
where the absorption features become heavily saturated
(for \op absorption at a overdensity of about ~3 at $z=3$. 
Without peculiar velocities  our inversion procedure works 
as well as we could reasonably expect. The obvious way 
to extend the inversion procedure to regions of higher 
density would be to use the higher order Lyman series lines 
which have smaller effective absorption cross sections and therefore 
saturate at increasingly higher densities (Cowie,  private communication). 
However, as we discuss below peculiar velocities significantly affect the
quality of the recovered density field, especially in high density
regions. The incorporation of higher order Lyman series lines in the 
inversion procedure will therefore be difficult and we leave this to 
future work. 

The second panel of Fig.~2 shows the DM density field recovered from 
a mock spectrum generated with peculiar velocities but before
correcting for redshift distortions.  
As expected the correspondence between true and recovered 
density is  degraded compared to  the case where the spectrum
was generated with the peculiar velocities set to zero. The main 
features of the density field are still recovered but there is a 
significant shift  between true and recovered density in real space 
and there of course remains the systematic underestimate of the
density in saturated regions. 

The third panel shows the reconstructed density for 
our full iterative procedure including the correction of 
 redshift distortions  as described in section 2.4.  
The shift between true  and recovered density in real space is reduced and the 
overall correspondence has significantly improved. This indicates that
we succeed in removing a major fraction of the redshift space 
distortions. The recovered and true l.o.s. velocity fields are shown in 
the bottom panel.  In all cases shown we have used the correct value 
for \cA, i.e. the value used to generate the mock spectrum. It should 
have become clear in this section that peculiar 
velocities significantly influence the quality of the recovery.

\subsection{The density probability distribution}  

For a statistical analysis of the recovered density field we will use
the probability distribution function  in differential (PDF) and cumulative
form (CPDF).   Let us for the time being still assume that the 
true value of \cA\ is known.  We will
discuss how \cA\ can be estimated from the PDF of the recovered
density field in the next section.  

Fig. 3  shows the CPDF for the true and recovered DM density field 
for the simulations  at $z=3$ and $z=0$. At low densities both curves
correspond very well but the  deviations become large  at high
densities due to saturation effects which confirms the visual
impression from  figure (\ref{flux:sim}) and (\ref{los:sim}).  
We further quantify the differences between the CPDFs 
of true and recovered density field in terms of
a number of moments  of the  density distribution in table \ref{tab1}. 
The bias in the recovered density introduces large discrepancies
between the  estimated and true values  of the moments. 
For example, the mean and rms values estimated at $z=0$ are: 
0.614 and 0.608; instead of the true values: $1$ and $3.52$.

\begin{figure}
\centering
\mbox{\psfig{figure=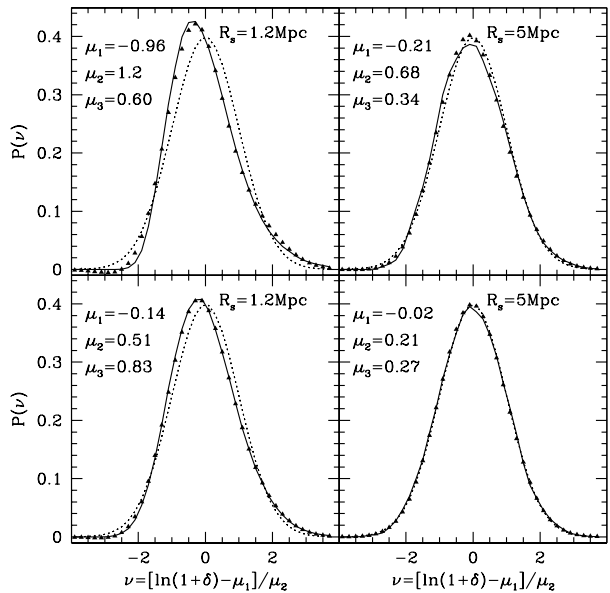,height=3.3in}}
\caption{The probability distribution  
of  DM density in the simulation at 
$z=0$ (top panels) and $z=3$ (bottom panels) 
in terms of $\nu= [ln(1+\delta)-\mu_1]/\mu_2$ 
(the density field was smoothed with a Gaussian windows of width 1.2Mpc and
5Mpc in comoving units, respectively).  The moments 
$\mu_1=<x>$, $\mu_2=<(x-\mu_1)^2>^{1/2}$ and $\mu_3=<(x-\mu_1)^3>/{\mu_2}^2$, 
where $x= ln(1+\delta)$, are indicated on the plot. The dotted 
curves are the best-fitting log-normal distribution, while 
the triangles show the best-fitting  Edgeworth  expansion of the PDF 
as in equation (\ref{edge}).}
\label{pdf:sim}
\end{figure}

What we would like to have is a simple parametric form for the PDF of 
the DM density. This could then be used to correct for the
biases in the recovered density field.
It has been suggested  (e.g. Kofman et. al. 1994, Coles \&  Jone 1991.) that
the PDF of the density field of a gravitationally clustering 
Gaussian random field in the  mildly non-linear regime 
is well described by a {\it log-normal} PDF. 
In that case  the quantity 
\begin{equation}
\nu \equiv [{\rm ln}(1+\delta)-\mu_1]/\mu_2  \label{nu}
\end{equation}
has  a normal (Gaussian) PDF, 
where $\mu_1$ and $\mu_2$ are the average and $rms$ values of ${\rm
ln}(1+\delta)$. In figure (\ref{pdf:sim}) we
compare the PDF of the DM density in the simulations (filtered with 
Gaussian windows of widths $R_s=$1.2Mpc and $R_s=$5Mpc in comoving 
units for $z=0$ and $z=3$)  with a log-normal distribution.  
For large smoothing scale the density field is indeed adequately 
described by a log-normal distribution.  However, for 
decreasing  smoothing scale, the true PDF becomes more 
and more skewed.  
The skewness of $\nu$  which we define as 
$\mu_3=<\nu^3/\mu_2>$ is one way of quantifying the deviation 
from a log-normal distribution and is also indicated on the plot.   

\begin{table}
\caption{Moments of the density field.  
The symbols $\langle . \rangle$ and $|.|$ 
denote average and absolute values respectively. 
Also $x=\delta/\sigma$, where
$\delta=\rho/<\rho>-1$. The {\it top} figures are for the 
true density filed, the {\it middle} figures are 
for the recovered density field and the {\it bottom} figures 
are for the best-fitting Edgeworth expansion as described 
in the text. The moments are computed assuming the true value of 
$\cal A$.}  
\begin{tabular}{|l|c|c|c|c|c} \hline
&\multicolumn{1}{c}{$\rho /\langle \rho \rangle$}&\multicolumn{1}{c}{
$\sigma$}
&\multicolumn{1}{c}
{$S_3=\langle \!x^3\!\rangle\!/\!\sigma$}
&\multicolumn{1}{c}{$S_{3r}=\langle
\!x\!|x|\!\rangle$}&\multicolumn{1}{c}{$S_{4r}
=\langle \!|x|\!\rangle $} \\ \hline 
True &&&&& \\
   z=3: & 1 &0.89 &4.85&0.66&0.61 \\ 
   z=0: & 1 &3.52  &5.08&0.93&0.33  \\ \hline 
Rec&&&&& \\
   z=3:& 0.80&0.51 &2.25&0.36&0.79 \\
   z=0:&0.61 &0.61 &1.97&0.40&0.83\\ \hline
Fit&&&&& \\
   z=3:&1     &1.12&4.27 &0.69&0.59\\
   z=0:&1     &2.97 &8.75&0.85&0.37\\ \hline 
\end{tabular}
\label{tab1}
\end{table}

To obtain an improved  fit to the  
PDFs of our simulation we use the first term of 
an Edgeworth  expansion (Colombi 1994, Juszkiewicz et. al. 1995),
\begin{equation}
P(\nu)= G(\nu)\left[
1+\frac{1}{3!}{\mu_3 \mu_2}(\nu^3- 3\nu)\right] ,
\label{edge}
\end{equation}
where $G(\nu)$ is a Gaussian with zero mean and unit variance.  
By imposing the condition $<\delta>=0$ we find the 
following relation between the three parameters,
\begin{equation}
\mu_1 + \frac{\mu_2^2}{2}+{\rm ln}\left(1+\frac{\mu_2^4
\mu_3}{3!}\right)=0 . \label{mean}
\end{equation} 
The functional form (\ref{edge}) has  therefore 
two free parameter (one more than a log-normal PDF). 
In the remainder of this section we arbitrarily choose 
to work with $\mu_2$ and $\mu_3$.  We also tried to  reduce 
(\ref{edge}) to a one-parameter family. 
We failed, however, at establishing a tight relation
between the two parameters for the PDFs of our simulations, 
which holds at all redshifts (see also Colombi 1994). 

In practice we find it  again more convenient to work with the 
CPDF rather than the PDF itself,
\begin{eqnarray}
Q(\rho)&\equiv& \int_{\rho}^{\infty}{P({\nu})\dd \nu}  \nonumber \\
&=&
\frac{1}{2}{\rm erfc}\left(\frac{\nu}{\sqrt{2}}\right)+
\frac{\mu_2\mu_3}{3! \sqrt{2\pi}}
\left(\nu^2-1\right)\exp\left(-\frac{\nu^2}{2}\right) . \nonumber \\
\label{cpdf}
\end{eqnarray}
We  fit the functional form (\ref{cpdf}) for the CPDF 
with $\mu_1$ expressed in terms of $\mu_2$ and $\mu_3$
(\ref{mean}) to  $Q_0(\rho)$ (the CPDF computed directly 
from  the recovered density)  by  
minimizing the quantity  
\begin{equation}
R=\int_0^{\rho_{\rm max}} \dd \rho
\left[Q(\mu_2,\mu_3;\rho)-Q_0(\rho)\right]^2 . \label{resid}
\end{equation}
with respect to $\mu_2$ and $\mu_3$. The cutoff $\rho_{max}$ is 
introduced in order to give no weight to 
high-density regions where the discrepancy between 
the recovered and true densities is severe.
Fig. \ref{logfit}  and Fig.  \ref{pdf:sim}  show these fits 
as triangles.  The quality of the fit is generally satisfying.
At the bottom of table \ref{tab1}  we list the corresponding  moments. 
The mean is correctly recovered by construction but the other moments 
are also in  significantly better agreement  with the moments   
of the true density field  than those calculated directly from the
recovered density field. The rms value at $z=0$ is now e.g. 
$2.97$ as compared to $0.62$ from direct calculation 
and $3.52$ for the true density field. When we  fitted a log-normal 
distribution to the  CPDF we found similar values for 
the moments as for the Edgeworth expansion.

\subsection{Determining ${\cal A}$ ($ \propto \Omega_{\rm bar}^2/J$)}

So far we have assumed that we know the  value of \cA\ . As is clear 
from  equation (\ref{nhi}) and  (\ref{tauf}) assuming a wrong value of  
\cA\ leads to an estimate of  $\rho_{DM}/\bar \rho_{\rm DM}$ 
which  is wrong by a constant factor ${(\cal A/ \cal A_{\rm
true})}^{1/\alpha}$ independent of the density. 
It is therefore possible to directly
estimate \cA\  by fitting  one of the parametric forms discussed in
the last section to the CPDF of the recovered density with \cA\ 
being left as an additional free  parameter. It turned out that the Edgworth 
expansion is less suitable for this procedure as is already has two 
free parameters (it becomes then sometimes difficult to get a unique fit). 
Fig. 5  shows the residuals of such a fit for the log-normal 
distribution as a function of  \cA\  at the two redshifts. The
residuals change significantly with varying \cA\ and show  pronounced
minima but the values  of \cA\ at the minima are biased low  
by about 25 percent and 15  percent, respectively. The error of \cA\ 
determined in this way is probably of the same order as when  
\cA\ is estimated by adjusting the mean flux level in mock 
spectra generated from numerical simulations. 
We also show the case for a recovery from mock spectra 
with the peculiar velocities set to zero (dotted curves).  
The minima shift significantly. This  demonstrates 
that the bias in \cA\ is not only  due to the systematic 
underestimate of the density in high-density regions and the 
inaccuracy of the log-normal distribution but also 
due to the  effect of peculiar velocities. Peculiar velocities  
introduce a  dependence  of the bias on the amount of
small-scale power  in the density fluctuation spectrum as apparent 
from the comparison between the cases at $z=3$ and $z=0$ 
which differ by a factor four in the rms density fluctuations. 
More numerical work is needed to  quantify  the effect of 
changing the cosmological model/power spectrum and the resolution, but
it seems nevertheless feasible to correct for the bias in \cA\ to 
about 10 percent accuracy.  

\begin{table}
\caption{Sensitivity of the estimated \cA\ and the moments 
to uncertainties in $\alpha$.The bottom line shows the effect of varying 
the mean flux from $0.69$ (as in table 1) to $0.75$ 
($\alpha=1.7$ was assumed).Results  are for  $z=0$. }
\begin{tabular}{|l|l|c|c|c|c|c} \hline
&\multicolumn{1}{c}{${\cal A}_{\rm min}$}&\multicolumn{1}{c}{$\mu$}&\multicolumn{1}{c}{
$\sigma$}
&\multicolumn{1}{c}
{$S_3$}
&\multicolumn{1}{c}{$S_{3r}$}&\multicolumn{1}{c}{$S_{4r}$} \\ \hline
 $\alpha=2.0$ &0.72 &1 &  2.00&6.35 & 0.82&0.46\\ \hline
 $ \alpha=1.7$ &0.84&1 &  2.97&8.75 & 0.85&0.37\\ \hline
 $ \alpha=1.5$ &1.08 &1 &  3.70&15.64 & 0.82&0.31\\ \hline
 $\langle \!F\!\rangle =0.745$ &1&0.92 &2.89 &8.89&0.85&0.37 \\ \hline
\end{tabular}
\label{tab2}
\end{table}

\begin{figure}
\centering
\mbox{\psfig{figure=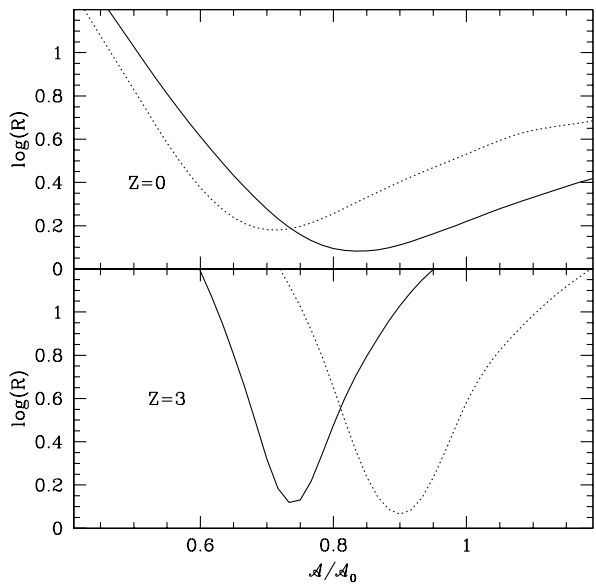,height=3.2in}}
\caption{Residuals of the fit of a  log-normal distribution to the 
distribution function of the recovered density as a function of 
the ratio of assumed to true value of ${\cal A}$ . 
Solid and dotted lines refer to mock spectra generated 
with and without peculiar velocities, respectively.}
\label{table}
\end{figure}

\begin{figure}
\centering
\mbox{\psfig{figure=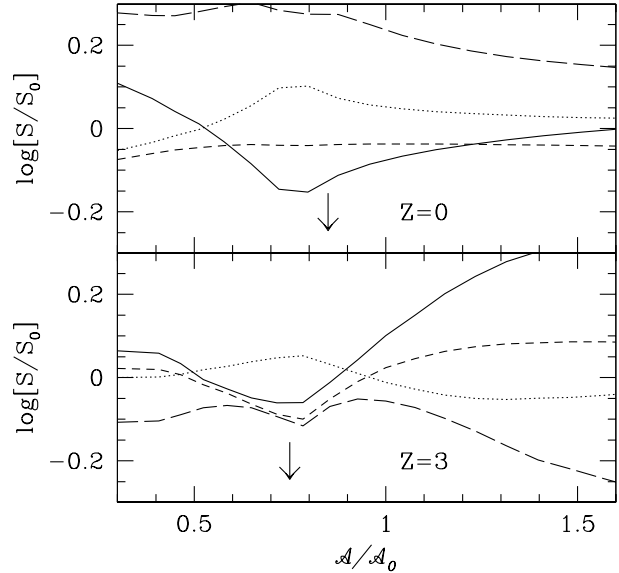,height=3.2in}}
\caption{Moments of the density probability distribution 
determined by fitting an Edgeworth expansion as a function 
of the assumed value of \cA\ . All values are divided by their true values.
Solid, long-dashed, short-dashed and dotted lines correspond to 
$\sigma$, $S_3$, $S_{3r}$ and $S_{4r}$ as defined in table 1. 
The arrows indicate the position of the minima in 
Fig. 5 for the case with peculiar velocities. }
\label{table}
\end{figure}

The assumed value of \cA\ will obviously also affect the determination
of the moments described in the last section. 
Figure 6 shows the dependence of the estimated moments on the assumed 
value of \cA. For a reasonable range of \cA\ the biases range from 
20 to 50 percent. 
If  \cA\ is determined by fitting a 
log-normal CPDF and the moments are then estimated by fitting 
an Edgeworth expansion to the CPDF with this value of \cA\ the 
biases in the moments are generally smaller than 25 percent.

We have  also  run our inversion procedure with equation of states  
different from the one used to generate the mock spectra and changed  
the mean flux  in order to test how sensitive the biases are  
to these parameters. The corresponding estimates of \cA\ 
and the moments are  listed in table 2. At least for the parameter
space explored the biases seem  to be robust.

\section{Discussion and Conclusions}

We have used numerical simulations to demonstrate  that the l.o.s.  
matter density field  can be recovered from QSO absorption spectra 
using an analytical model for the IGM and a Lucy-type iterative 
inversion algorithm. We have thereby estimated the l.o.s. peculiar 
velocity field from the recovered density field and have corrected 
for redshift distortions in an iterative procedure. The inversion 
works well for the most underdense regions up to  the density where the
absorption features saturate. For \op this occurs at an overdensity of
a few but this limit can  be pushed to higher densities by 
incorporating higher order lines of the Lyman series. , 
However, an inversion will become increasingly more difficult
at higher densities due to shell crossing and shock heating
of the gas.    

We have then used the fact that the density probability distribution 
seems to have a universal shape to get reliable estimates 
of higher order moments of the PDF  despite the missing high-density 
tail. We found that the Edgeworth expansion is an excellent 
approximation to the PDF of the logarithm of the density, 
consistent with the results of Colombi (1994).  By fitting the first
two terms of such an Edgeworth expansion we  estimated  a number of
moments of the PDF with an accuracy of about 25 percent. Estimates of similar 
accuracy were obtained for the parameter combination \norm by fitting 
to a log-normal distribution. We expect that, eventually,
these biases can be corrected to 10 percent accuracy. 

In principle it should also be possible to estimate the correlation 
function and the power spectrum directly from the recovered density
field. These will, however,  be affected by peculiar motions, 
the inability to recover high density regions and the  
bias in the determination of \cA\ . We have not yet investigated in 
detail how big the  resulting errors are.  
It might well be that a  comparison of observed flux power spectra 
with those of mock spectra as suggested by Croft et al. is the most 
favourable way to deal with these problems.

Observational information on the PDF of the matter density 
comes so far mainly from  the analysis of galaxy surveys and is
restricted to  small redshifts. These constraints could be checked
and extended to a much wider redshift range by applying our inversion  
technique to a moderate number of QSO absorption spectra of the kind 
which are now routinely taken with  10m class telescopes. This should 
also check and tighten existing constraints on the evolution of the 
UV background. By combining our inversion technique with a redshift 
survey along QSO sightlines the clustering of galaxies and matter can  
be related without referring to a particular cosmological model.  Such a
determination of the bias between galaxy and matter clustering 
is especially worthwhile. Most currently favoured models agree 
with the recently determined  clustering strength of high-redshift 
galaxies but differ strongly in their predictions for the 
yet unknown bias relation (Adelberger et al. 1998). 
Three-dimensional information on the density and peculiar velocity 
field can be gained by applying the method  to two or  more spatially 
close line-of-sights.  An intermediate-resolution spectral survey of 
quasars down to about 22nd magnitude in a field of a couple of square 
degree size has e.g. been suggested as a possible project for the VLT 
(Petitjean 1996). Such a survey will result in about 100 l.o.s. in a
region about 30Mpc across. With the method proposed in the appendix
such a region could become a unique laboratory for the study of how
galaxy formation is related to the distribution and dynamics of the 
underlying matter field.

\section{Acknowledgments} 
We thank Rupert Croft, Michael Rauch, Ravi Sheth, 
David Weinberg and  Simon White 
for helpful comments. AN thanks Yehuda Hoffman for many 
discussions on constrained realisations.  
This work was supported
in part 
the EC TMR network for ``galaxy formation and evolution''
and the ``Sonderforschungsbereich 375-95 f\"ur Astro-Teilchenphysik der 
Deutschen  Forschungsgemeinschaft''.

\protect

\appendix

\section{Peculiar velocities  and 3-D densities 
from line-of-sight density fields}

\subsection{From l.o.s. density to l.o.s. peculiar velocity}
In this section we describe the method which we used to estimate  the 
l.o.s. peculiar velocity component from the 
(estimated)  {\it real} space density along the l.o.s.    We first have to
assume a non-linear relation between the 3-dimensional velocity and
density fields. Various such relations have been suggested, here
we work with the relation found empirically by Nusser et. al. (1991).
Let $\th = -{\rm div}\vv$, where $\vv$ is the physical peculiar
velocity and the divergence is with respect to the physical coordinate
$r$ in units of ${\rm km/s}$. The approximation suggested by 
Nusser et al.~(1991) is given by 
\begin{equation}
\th = \frac{\delta}{1.+0.18\delta} +const . \label{nus}
\end{equation}
The relation is local and provides $\th$ in terms of the density
contrast along the l.o.s.   The constant factor is introduced 
to ensure $<\th>=0$. Since $\th$ involves derivatives of velocity 
components perpendicular to the l.o.s. , the l.o.s.  peculiar velocity
component can not uniquely be determined from $\th$. Therefore, the
best we can do is provide an estimate for the l.o.s. peculiar velocity
based on some statistical assumptions. Neither of the fields $\th$ or
$\delta$ is Gaussian. However, for an approximate treatment we assume
here that $\th$ is Gaussian. Note that by (\ref{nus}) this assumption
does not imply Gaussian density fluctuations. At the end of the next
subsection we will briefly describe how estimates of the peculiar
velocity can be obtained without assuming that the field $\th$ is
Gaussian.  Let $\cv$ be the component of $\vv$ along the l.o.s. 
Subsequently we work with the (1-dimensional) Fourier tranforms,
$\tht(q)$ and $\cvt(q)$ of the l.o.s. fields $\cv$ and $\th$.   
According to Bi (1993) $\tht$ and $\cvt$ are related by
\begin{equation}
\tht(q)=\ut(q)+\wt(q) , \label{bi1}
\end{equation}
\begin{equation}
\cv(q)=iq\alpha(q)\wt(q) , \label{bi2}
\end{equation}
where  $\wt(q)$ and $\ut(q)$ are two uncorrelated Gaussian
fields with power spectra ${\rm P}_w$ and ${\rm P}_u$.
These power spectra can be written, in terms  of the 3D matter 
power spectrum $\rm P$,

\begin{equation}
{\rm P}_w(q)=2\pi \alpha^{-1}\int_q^{\infty}{\rm P}(k)k^{-1} \dd k ,
\end{equation}
and
\begin{equation}
{\rm P}_u(q)=2\pi\int_q^{\infty}{\rm P}(k)k\dd k - {\rm P}_w(q) ,
\end{equation}
with
\begin{equation}
\alpha(q)=\frac{\int_q^{\infty}{\rm P}(k)k^{-3} d k}{
\int_q^{\infty}{\rm P}(k)k^{-1} \dd k}.
\end{equation}

The relations (\ref{bi1}) and (\ref{bi2}) can be used to 
estimate  $\cvt(q)$ from $\tht(q)$.  For simplicity, 
we work with $\wt$ instead of $\cvt$ and switch back to $\cvt$
at the end of the calculation.  According to Bayes' theorem the
conditional probability ${\rm P_r}[w|\tht]$  is given by
\begin{equation}
{\rm P_r}(w|\tht)=\frac{{\rm P}(w)}{{\rm P}(\tht)}{\rm P}(\tht|w)
\end{equation}
Using the relations (\ref{bi1}) and (\ref{bi2}) we find
\begin{equation}
{\rm P_r}(\tht|w)\propto  \exp\left[-\frac{(\tht-w)^2}{2{\rm P}_u}\right] ,
\end{equation}
and
\begin{equation}
{\rm P_r}(\wt)\propto \exp\left[-\frac{w^2}{2{\rm P}_w}\right] .
\end{equation}
Therefore
\begin{equation}
{\rm P_r}(\wt|\tht)\propto \exp\left[-\frac{(\tht-\wt)^2}{2{\rm
P}_u}-\frac{\wt^2}{2{\rm P}_w}\right]
\end{equation}
This is a Gaussian with mean
\begin{equation}
<\wt|\tht>=\left(1+\frac{{\rm P}_u}{{\rm P}_w}\right)^{-1} \tht ,
\label{wiener1}
\end{equation}
and variance (power spectrum)
\begin{equation}
{\rm P}_{w|\tht}=\frac{ {{\rm P}_w}^2}{{\rm P}_u+{\rm P}_w} .
\label{varwth}
\end{equation}
Thus, given $\tht(q)$, one may write
\begin{equation}
\wt(q)=\left(1+\frac{{\rm P}_u}{{\rm P}_w}\right)^{-1} \tht
+n(q) , \label{wienern}
\end{equation}
where $n(q)$ is a random Gaussian variable with power spectrum
$P_{\wt|\tht}$ which is uncorrelated with $\tht(q)$.  The form
of (\ref{wiener1} is reminiscent of Wiener filtering of noisy data 
(Wiener 1949, see also Zaroubi et. al. (1995) and Fisher et. al. (1995) for 
applications to cosmology). To see this simply identify  ${\rm P}_u$ and 
${\rm P}_w$ as, respectively,  the signal and noise power spectra.
According to (\ref{bi1}), the power spectrum of the unconditional 
$\tht$ is ${\rm P}_{\theta}= {\rm P}_w+{\rm P_u}$. This clearly
differs from the power spectrum of $ <\wt|\tht>$ given
in  (\ref{varwth}). 
Following Sigad et. al. (1998) here we  replace the filter 
$(1+{\rm P_u}/{\rm P_w})^{-1}$ in (\ref{wiener1}) by  $(1+{\rm P_u}/{\rm
P_w})^{-1/2}$ in order to preserve the power spectrum of the
unconditional field. 
Alternatively we could have complemented $<\wt|\tht>$ with the random
field $ n(q)$ in order to preserve the power.

\subsection{Reconstruction of 3D fields}

For simplicity we discuss the case of a single l.o.s.  The
generalisation to multiple parallel lines of sight is obvious.
We arbitrarily choose the l.o.s. to be along the $x$ axis.
Once the density $\dl$ in the l.o.s. is given, we can easily compute 
the corresponding Fourier transform $\dtl (q)$ defined by
\begin{equation}
\dtl (q)=\frac{1}{(2\pi)^{1/2}}\int \dl (x)\exp(i qx)\dd x 
\label{ft1d}
\end{equation}
We write the 3D density field $\d (\vr)$ in terms of its Fourier transform 
\begin{equation}
\d (\vr)=\frac{1}{(2\pi)^{3/2}}\int \dt (\vk) \exp(-i \vk\cdot
\vr)\dd^3 \vk . \label{ft3d}
\end{equation}

Define $l(x)=\d(\vr=x {\hat x}) $ where $\hat x$ is a unit vector
along the $x-$axis. Hence, by combining (\ref{ft1d}) and (\ref{ft3d})
we obtain
\begin{eqnarray}
\dtl (q)&=&\frac{1}{(2\pi)}\int \dt (\kpar,\kper) 
\exp(iqx-i\kpar x) \dd \kpar \dd^2 \kper \dd x \nonumber \\
&&\nonumber \\
&=&\int \dt (q,\kper)  \dd^2 \kper  , \label{ft1d3d}
\end{eqnarray}
where $\kpar$ and $\kper$ are the components
parallel and perpendicular to the l.o.s.   The problem of
reconstructing the 3D density field reduces to estimating $\dt(\vk)$
given the coefficients $\dtl(q)$ and the relation (\ref{ft1d3d}). Note
that $\dtl$ can be computed from $\dl$. 
The method of Hoffman and Riback (1991) (hereafter HR)
can be used to obtain such an estimate. First, we 
derive an expression for the mean value 
$\dt^{\rm MV}(\vk)=<\dt(\vk)|\{\dtl(q)\}> $
of  $\dt(\vk)$ corresponding to a particular $\vk$ 
given the set of coefficients $\{\dtl(q)\}$.
\def\qp{q^\prime}
According to HR, we have   
\begin{equation}
\dt^{\rm MV}(\vk)=\int M(\vk,\qp) N^{-1}(q,\qp) \dtl(q)
\dd q\dd \qp ,\label{HFexp}
\end{equation}
where $M(\vk,\qp) =<\dt(\vk) \dtl(q)>$ is  the covariance matrix of $\dt$
and $\dtl$, and $N^{-1}(q,\qp)$ is the {\it inverse} of the
covariance matrix $N(q,\qp)=<\dtl(q) \dtl(\qp)>$; both matrices are of
infinite dimension.  
Using the homogeneity condition,
$<\dtl(\vk)\dtl(vk^{\prime}>=\delta^D(\vk-\vk^\prime){\rm P}(k)$ and the
relation (\ref{ft1d3d}) it can be seen that 
$N$ and $M$ are given by
\begin{equation}
N(q,\qp)=\delta^{\rm D}(q-\qp) {\rm P}^{\rm los}(q) ,
\end{equation}
and
\begin{equation}
M(\vk,q)=\delta^{\rm D}(q-\kpar){\rm P}\left(\sqrt{q^2+\kper^2}\right)
\end{equation}
where  $\delta^{\rm D}$ is the Dirac $\delta-$function and
\begin{equation}
{\rm P}^{\rm los}=2\pi \int_q^{\infty} {\rm P}(k) k \dd k 
\end{equation}
is the 1D power spectrum  of the density field along  the l.o.s. 
The mean is given by 
\begin{equation}
\dt^{\rm MV}(\vk)=
\frac{{\rm P}(k)}{{\rm P}^{\rm los}(\kpar)}\;\dtl(\kpar) . 
\label{MV}
\end{equation}

Once the mean values are given, a constrained random realisation, 
${\tilde
\Delta}^{\rm C}$ can be generated from an unconstrained random
realisation ${\tilde \Delta}$ using (HR), 
\begin{equation}
{\tilde \Delta}^{\rm C}(\vk)=
\tilde \Delta(\vk)+\frac{{\rm P}(k)}{{\rm P}^{\rm los}(\kpar)}\;
\left[\dtl(\kpar)- {\tilde \Delta}^{\rm los}(\kpar)\right] ,
\label{CR}
\end{equation}
where ${\tilde \Delta}^{\rm los}(\kpar)$ are the
1D Fourier coefficients of the unconstrained  density field in the
l.o.s. 

The treatment so far relied on the HR method which assumes Gaussian
fields. This treatment will not be satisfactory for 
a 3D reconstruction in the  non-linear regime. The
unconstrained density field as given by (\ref{CR}) is e.g. not guaranteed
to have positive values everywhere. The scheme can be extended 
to the non-linear regime by two modifications: 
(i) use ${\rm ln}(1+\delta)$ instead of $\delta$,
(ii) extract an  unconstrained random field $\Delta(\vr)$ or its 
Fourier transform ${\tilde \Delta}(\vk)$ from a fully non-linear N-body
simulation.  The simulations can be used to  compute the 1D
and 3D power spectra of ${\rm ln}(1+\delta)$ which  are 
required to generate the constrained realisation. But note that these
power spectra  can in principle be estimated from the density field 
along the l.o.s. 

A 3D constrained realisation obtained with this scheme 
adapted to the  non-linear regime  can then be used
to estimate the peculiar velocity without assuming $\th$ 
to be Gaussian. $\th$ can be computed  at any point in space from 
the 3D constrained realization of the density field using  
the velocity-density relation (\ref{nus}) and  assuming a potential
flow.  The peculiar velocity is then obtained by solving  a Poisson-like 
equation to recover $\th$. 
The implementation of this alternative scheme for 
recovering the peculiar velocity field  and a comparisons to the
scheme described in the last section is left to future work.


\bigskip
\begin{thebibliography}{}
\def\ref{\par\noindent\hangindent 15pt} 
\def\apj#1{{ApJ,}
{ #1}} 
\def\apjl#1{{\it Astrophys. J. (Lett.)} { #1}} \def\apjs#1{{\it
Astrophys. J. (suppl.)} { #1}} \def\mn#1{{MNRAS,} { #1}}
\def\aa#1{{\it Astron. Astrophys.} { #1}} \def\nat#1{{\it Nature} {
#1}} \def\ana#1{{\it Annu. Rev. Astron. Astrophys.} { #1}}


\def\ref{\par\noindent\hangindent 15pt}
\def\apj#1{{\it Astrophys. J.} { #1}} \def\aj#1{{\it Astronomical. J.}
{ #1}} \def\apjl#1{{\it Astrophys. J. (Lett.)} { #1}} \def\apjs#1{{\it
Astrophys. J. (suppl.)} { #1}} \def\mn#1{{\it M.N.R.A.S.} { #1}}
\def\aa#1{{\it Astron. Astrophys.} { #1}} \def\nat#1{{\it Nature} {
#1}} \def\ana#1{{\it Annu. Rev. Astron. Astrophys.} { #1}}

\bibitem{} Adelberger K.L., Steidel C.S., Giavalisco M.,
Dickinson M.E., Pettini M., Kellog M., 1998, ApJ, in press, astro-ph/9804236

\bibitem[]{} Bahcall J.N., Salpeter E.E 1965, ApJ , 142, 1677

\bibitem{} Bechtold J., Crotts A.P.S., Duncan R.S., Fang Y., 
1994, ApJ, 437, L83

\bibitem{} Bi H.G., B\"orner G., Chu Y.  1992, A\&A, 266, 1 

\bibitem{} Bi H.G., 1993, ApJ, 405, 479

\bibitem{} Bi H.G., Davidsen A.F.,  1997, ApJ, 479, 523

\bibitem{} Bond J.R., Szalay A.S., Silk J., 1988, ApJ, 324, 627

\bibitem{} Cen R., Miralda-Escud\'e J., Ostriker J.P., Rauch M.,
1994, ApJ, 437, L9

\bibitem{} Coles, P., Jones, B.J.T, 1991, MNRAS, 248, 1

\bibitem{} Colombi, S., 1994, ApJ, 435, 536

\bibitem{} Cooke A.J., Espey B., Carswell R.F., 
1997, MNRAS, 284, 552



\bibitem{} Couchman H.M.P., Thomas P.A., Pearce F.R., 1995, ApJ, 452, 797

\bibitem{} Dinshaw N., Impey, C.D., Foltz C.B., Weymann R.J., Chaffee  F.H.,
1994, ApJ, 437, L87

\bibitem{fish} Fisher K.B., Lahav O., Hoffman Y., Lynden-Bell D., 
 Zaroubi S., 1995, MNRAS, 272, 885

\bibitem{} Gnedin N.,  Hui L., 1998, MNRAS, in press, astro-ph/9706219

\bibitem{} Gunn J.E., Peterson B.A., 1965, ApJ, 142, 1633

\bibitem{} Haehnelt M., Steinmetz M., 1998, MNRAS, in press, astro-ph/9706296  


\bibitem{} Hernquist L., Katz N., Weinberg D.H.,  Miralda-Escud\'e J.,
1996, ApJ, 457, L51

\bibitem{} Hui L., Gnedin N.Y., 1997, MNRAS, 292, 27

\bibitem{hf} Hoffman, Y.,  Ribak, E., 1991, ApJ, 380L, 5

\bibitem{} Jenkins A., Frenk, C.S., Pearce, F.R., Thomas, P.A.,
Colberg, J.M., S.D.M., Couchman, H.M.P., Peacock, J.A., Efstathiou,
G., Nelson, A.H., 1998, ApJ, in press, astro-ph/9709010.

 
\bibitem{L74} Lucy L. B., 1974, \aj{79}, 745

\bibitem{} MacFarland T.J., Couchman H.M.P., Pearce F.R., 
Pichlmeier J., 1998, astro-ph/9805096

\bibitem{} McGill C., 1990, MNRAS, 242, 544


\bibitem{} Miralda-Escud\'e J., Cen R., Ostriker J.P., Rauch M.,
1996, ApJ, 471, 582

\bibitem{N91} Nusser A., Dekel A.,  Bertschinger E., 
 Blumenthal G.,  {1991}, \apj{379} 6

\bibitem{} Petitjean P., M\"ucket  J.P., Kates R.E., 1995, A\&A, 295, L9

\bibitem{} Petitjean P., 1997, in: The Early Universe with 
the VLT, ed. Bergeron J., ESO Workshop, Springer, p.266

\bibitem{} Rauch M., Miralda-Escud\'e J., Sargent W.L.W., Barlow
T.A., Weinberg D.H., Hernquist H., Katz N., Cen R., Ostriker J.P.
1997, ApJ, 489, 7 

\bibitem{} Rauch M., 1998, ARAA, in press

\bibitem{} Sigad, Y., Eldar, A., Dekel, A., Strauss, M.A.,
Yahil, A., 1998, ApJ, 495, 516


\bibitem{} Wechsler, R.H., Gross, M.A.K., Primack, J.R., Blumenthal,
G.R., Dekel, A., 1998, ApJ, in press,  astro-ph/9712141

\bibitem{} Weinberg D.H,  1992,  MNRAS, 254, 315

\bibitem{} Weinberg D.H,  Miralda-Escud\'e J., Hernquist L., Katz,
N.,1997, ApJ, 490, 564

\bibitem{} Weinberg D.H,  Miralda-Escud\'e J., Hernquist L., Katz N.,  
Miralda-Escud\'e J., ,1998, in: Structure and Evolution 
of the IGM from QSO absorption lines, eds. Petitjean P., Charlot S.,
Editions Frontieres, p. 133 


\bibitem{wien} Wiener, N. 1994, in {\it Extrapolation and Smoothing of
Stationary Time Series}, (New York, Wiley)

\bibitem{} Yahil, A., Strauss, M.A., Davis, M., Huchra, J.P.,
1991, ApJ, 372, 380

\bibitem{zar} Zaroubi S., Hoffman Y., Fisher K.B.,  Lahav O., 
1995, ApJ,449, 446

\bibitem{} Zhang Y., Anninos P., Norman M.L., 1995, ApJ, 453, L57




\end{thebibliography}
\end{document}